\relax
\documentclass[letterpaper]{article} 
\usepackage{aaai21}  
\usepackage{times}  
\usepackage{helvet} 
\usepackage{courier}  
\usepackage[hyphens]{url}  
\usepackage{graphicx} 
\usepackage{natbib}  
\usepackage{caption} 
\usepackage{multirow}
\usepackage{adjustbox}
\usepackage{booktabs}
\usepackage{dcolumn}

\title{You Are How (and Where) You Search? Comparative Analysis of Web Search Behaviour Using Web Tracking Data}

\author {
    Aleksandra Urman,\textsuperscript{\rm 1} \textsuperscript{\rm 2}
    Mykola Makhortykh, \textsuperscript{\rm 1}\\
}
\affiliations {
    \textsuperscript{\rm 1} University of Bern \\
    \textsuperscript{\rm 2} University of Zurich\\
    urman@ifi.uzh.ch, mykola.makhortykh@ikmb.unibe.ch
}

\begin{document}

\maketitle

\begin{abstract}
We conduct a comparative analysis of desktop web search behaviour of users from Germany (n=558) and Switzerland (n=563) based on a combination of web tracking and survey data. We find that web search accounts for 13\% of all desktop browsing, with the share being higher in Switzerland than in Germany. We find that in over 50\% of cases users clicked on the first search result, with over 97\% of all clicks being made on the first page of search outputs. Most users rely on Google when conducting searches, and users’ preferences for other engines are related to their demographics. We also test relationships between user demographics and daily number of searches, average share of search activities in one’s general browsing behaviour as well as the tendency to click on higher- or lower-ranked results. We find differences in such relationships between the two countries that highlights the importance of comparative research in this domain. Further, we observe differences in the temporal patterns of web search use between women and men, marking the necessity of disaggregating data by gender in observational studies regarding online information behaviour.

\end{abstract}

\section{Introduction}
Web search engines are ubiquitous nowadays and act as major information gate-keepers in high-choice media environments. Google alone handled around 6.9 billion queries per day in 2020 \cite{petrov_most_2019} with an average user of Google.com turning to the site 18.15 times per day as of April 2021 \cite{alexa_googlecom_nodate}. When Google experienced an outage in 2013 for 5 minutes, there was a drop of 40\% in the global web traffic \cite{svetlik_google_nodate}. The numbers are staggering, especially given that Google is just one of the search engines - though the dominant one on most markets. Furthermore, search engines are highly trusted by their users:
according to Edelman Trust Barometer \cite{noauthor_2021_nodate}, in 2020 search engines were reported to be the most trusted information source globally. 

Given the importance of search engines for shaping public opinion, it is crucial to understand users' web search behaviours. Yet, our knowledge in this context remains limited and primarily relies on two types of data.
eye-tracking \cite{pan_google_2007,schultheis_we_2018} and  search engine transaction log data \cite{jansen_how_2006,weber_who_2011}. Both of these data sources have their limitations:
eye-tracking studies typically rely on small user samples and can hardly be generalized to broader populations.
On the contrary, log-based studies capture the behaviour of the large groups of users, but on the aggregate level, thus limiting possibilities for inferring the impact of users’ individual characteristics on how they search for information. Additionally, log-based studies can not 
reliably infer the connection between search results ranking and user behaviour, because researchers can not, in retrospect, identify how search results were ranked and presented to individual users due to the temporal changes in the results and effects of search personalization \cite{hannak_measuring_2013,kliman-silver_location_2015} and randomization \cite{makhortykh_how_2020,urman_matter_2021}.

In the present study, we address these limitations 
by relying on a type of data source that, to the best of our knowledge, has not been used in the context of web search behaviour. We utilize the combination of web tracking data \cite{christner_automated_2021} with demographic data about individual users acquired via survey to explore users' web search behaviour. Web tracking data includes information on user desktop-based browsing behaviour along with the actual HTMLs of the browsed content. By acquiring HTMLs of pages viewed by the users, 
we can infer the exact composition and ranking of web search results users were exposed to and, consequently, find out which of these results they clicked on. 

Using the combination of web tracking and survey data collected in Germany and Switzerland in spring 2020, we aim to address several gaps in the existing scholarship on web search behaviour. First, we scrutinize the effect of individual demographic characteristics on search behaviour using a large sample of users. Second, unlike earlier large-scale (i.e., log-based) search behaviour studies, which were focused on single-country populations (usually, the US), our study offers a comparative perspective and goes beyond the US context. Third, we examine the user clicking behavior in relation to web search results ranking 
in real-life conditions - in contrast to eye-tracking studies 
that typically rely on smaller samples and are carried out in lab settings.

Specifically, we address the following research questions: 1) how frequently do users with different demographic characteristics and socio-economic status use search engines?; 2) what are the temporal patterns of web search use and do they differ by demographics? 3) are there demographic or socio-economic status-based differences in the choice of specific search engines (i.e., Google/Bing/other)?; 4) how does the rank of a search result relate to the clicking behaviour of users with different demographics? We also examine country-level differences in relation to each of the four questions.

\section{Related work}
Studies on web search behaviour to date have relied on either of the two data source types: eye-tracking and search engine transaction log data.

\textit{Eye-tracking-based studies} 
are typically conducted on smaller samples, usually not demographically representative ones, and within lab settings. The advantage of such studies is that they allow examining user attention patterns in the context of web search and, for instance, exploring the relation between the ranking of search results and users' clicking behaviours. In one of the earliest studies \cite{granka_eye-tracking_2004}, the authors have examined attention and clicking patterns of web search engine users based on a student sample (n=36), and found that top-ranked results receive disproportionately more attention and clicks than lower-ranked ones. This finding was corroborated in numerous further studies (e.g., \cite{pan_google_2007,schultheis_we_2018,joachims_accurately_nodate,guan_eye_2007}). 

Eye-tracking studies have also investigated the impact of additional factors 
on search result selection. For instance, two studies used a small (n=18) sample of users of diverse ages and occupations \cite{joachims_accurately_nodate} and a student sample (n=22) \cite{pan_google_2007} from the US found that clicking decisions are influenced not only by ranking but also perceived relevance of search results. 
A replication of the latter study \cite{schultheis_we_2018} conducted circa 10 years after the original one on a student sample (n=28) in Germany has found similar effects thus 
indicating the stability of observed effects across time and different national contexts. 

Despite providing important insights in user search behaviour, eye-tracking studies are subjected to a number of limitations, in particular their limited scalability. While some 
potential solutions for scaling are being offered in recent years - e.g., eye-tracking via webcam devices \cite{papoutsaki_searchgazer_2017} - their precision remains 
lower than that of more conventional lab-based dedicated eye-trackers \cite{holmqvist_eye_2011}. Due to the scalability problem, eye-tracking studies are based on small samples which are not demographically representative and, often, are made of student samples recruited in the US. This leads to a limited generalizability of eye-tracking-based findings 
: it is unclear whether users with different demographics search the web in similar ways, and whether there are country-level differences in how they do it.

\textit{Transaction logs-based studies}, on the contrary, allow examining web search behaviour on a large scale. One of the earliest studies utilizing transaction log data \cite{silverstein_analysis_1999} was conducted in 1999. Based on circa 1 billion search queries entered into AltaVista search engine over a period of 6 weeks the authors found that users tend to type in short queries and rarely navigate beyond page 1 of the search engine. Similar findings were reported by authors of a 1-week-long study based on a Korean search engine Naver \cite{park_end_2005}. 

Log data has also been utilized to examine temporal aspects of web search \cite{zhang_time_2009} and the patters of search query usage \cite{weber_who_2011}. Such studies allow inferring real-life web search usage patterns and are based on large data samples - as contrasted to eye-tracking-based lab studies. However, log-based studies 
also have several limitations. First, 
due to the difficulty of obtaining search logs data 
owned by proprietary companies, most of the transaction logs-based studies focus on single search engines. It undermines
the generalizability of their findings since usage patterns o
can be affected by the differences in search engine interfaces and/or the differences in the demographics of their users. Even studies such as \cite{jansen_how_2006} that analyze log data from multiple search engines can not match the users across these engines, which prevents them from examining if
the same users utilize multiple different engines and, if so, whether and how their behavior is different depending on the engine. 

The absence of reliable demographic data about the users is another 
limitation of logs-based studies. Such data is sometimes available on users' gender and age, but not on other variables
such as education or income level that can only be inferred by the researchers 
(e.g., \cite{weber_who_2011}). However, even such inference-based studies 
are rare 
Third, log-based data is inherently noisy, because search requests might be executed not only by human users but also by bots, and it is difficult to differentiate between organic and automated requests \cite{jiang_mining_2013}. Finally, transaction log data, 
does not allow tracing the position of the search results a user clicked on and the only ranking-related parameter available
is the number of the search result page on which a user selected a result.

The aforementioned limitations of both approaches can be addressed by utilizing web tracking data that includes full HTMLs of the pages browsed by users and is combined with survey data. Unlike eye-tracking, this approach is scalable and allows observing user behavior in real-life circumstances, not in a lab setting. Unlike with transaction logs, with tracking data it is possible to reliably know users' demographics (from the survey), observe user behavior across multiple search engines, make sure that the data comes from real users and not bots or machines and, finally, through matching the data from the scraped web search HTML with the URLs subsequently clicked by the user, infer the exact position of the result a user clicked on. Thus, this approach allows combining the strengths of both approaches previously utilized to measure web search behavior, while overcoming their limitations.

\section{Data and Methods}
\subsection{Data}
To collect data for our study, we recruited a sample of Internet-using participants in the age range of 18-75 years from Germany and German-speaking Switzerland. The recruitment was conducted via the market research company Demoscope in early March 2020 using online access panels with 200,000 members (Germany) and 35,000 members (Switzerland). Participants were randomly selected in accordance to quotas regarding gender, age, and education to construct a representative sample of the German and the German-speaking Swiss population. For Germany, the region of residence (West vs. East) was used as an additional sampling criterion.

The selected participants were invited to participate in the survey, which was completed by 1,952 participants in Germany and 1,297 in Switzerland. As a requirement to take part in the survey, participants were asked whether they agreed to participate in the online tracking study using a browser extension that records their online behavior. While agreement to be tracked was required to partake in the survey, participants were informed that they could opt out from being tracked at any time.

After agreeing to participate in online tracking, each participant received a link to the website where extensions (i.e., plugins) for desktop versions of Chrome and Firefox browsers could be downloaded and installed. The extensions were designed specifically for the project and based on the screen-scraping principle, namely capturing HTMLs of web content appearing in the browser, where the extension was installed \citep{christner_automated_2021}. The captured HTML content together with the URL address of the page from which it was captured were sent to the remote server, where data were encrypted and stored.

To protect participant privacy, the extensions were supplemented by a "hard" denylist (i.e. a list of websites whose content was not captured and visits to which were not recorded; this included insurance companies, medical services, pornography websites, bank websites, messengers, and e-mail services) and a "soft" denylist (i.e. a list of websites whose content was not captured, but the visits to which were recorded; the list included commercial websites). Participants were also provided with the possibility to switch browser extensions 'private mode', where no HTML content was captured, so they could browse privately if they felt the need to. 

Out of the original sample of participants expressing agreement to being tracked, 587 (Germany) and 601 (Switzerland) participants had successfully registered at least one website visit by the end of the tracking period (March 17 to May 26 2020). The present sample consists only of those who registered at least one web search - 563 participants in Switzerland and 558 in Germany. The participants were asked about their age, gender, education and income. The reported levels of education were collapsed into three subgroups to ensure comparability between the two countries: obligatory school only, full secondary education, tertiary education. Participants also reported their monthly income (according to pre-defined income breaks, different for Switzerland and Germany due to the differences in the overall income levels between the two countries). The demographic distributions in the samples are as follows: self-reported gender: CH - 43.8\% female, DE - 44.8\% female; age: CH - mean=43.8, median=42, DE - mean=49.3, median=51; education: CH - 3.6\% obligatory education, 54.9\% - full secondary education, 41.6\% - tertiary education, for DE corresponding numbers are 12.9\%, 51.3\%, 35.8\%; income: CH - 7.2\% not reported, 37.7\% below 3999CHF, 37.1\% between 4000 and 6999 CHF, 17.9\% above 7000 CHF; DE - 3.1\% not reported, 48.4\% below 1999 EUR, 44.8\% from 2000 to 4999 EUR, 3.8\% above 5000 EUR.

\subsection{Methods}
To filter out only web search visits from the overall tracking data, we used url-based filtering by domain first. This step was based on a list of search domains constructed by us based on the lists of engines commonly utilized by European users as indicated by sources such as \cite{noauthor_2021_nodate} and included the following engines: Google, AOL, Bing, DuckDuckGo (DDG), Ecosia, Gigablast, Metager, Qwant, Swisscows, Yahoo, Yandex. Then, we also filtered the data by subdomains and URL parts that would point to a service from a search engine company other than web search 
(e.g., Google Photos in the case of Google). Then, we calculated the share of visits to image search, video search as well as news search among all search traffic. In total, we have recorded 348018 user visits to text, image and video search across all engines combined. The results have demonstrated that visits to these services are infrequent: image search accounted for 0.2\% of search engine traffic, video search for 0.06\%. Thus, we focused on text search only as it accounted for over 99.7\% of all search traffic.

After filtering out 
text search results 
we merged the tracking data about participants' web search visits with their demographic data (self-reported gender and age) and socio-economic data (self-reported income and level of education) obtained via the survey. This merged data was used in the next steps of the analysis. Each step was performed separately for the German and the Swiss subsamples, with comparisons drawn between the two when reporting the results.

\textit{RQ1: frequency of search use.} To establish how frequently users with different demographic characteristics utilize
web search, we 
computed descriptive statistics about the average proportion of visits to web search engines to the overall number of visits tracked and average number of search queries executed daily by users from different demographic (age and gender) groups. Then, we tested the association between user demographics and socio-economic characteristics, and frequency of their web search usage via
a generalized linear model. We used
the average number of searches executed by each user as a dependent variable, and users' characteristics as predictors. In this and other regression models described below we controlled for the users' overall web activity as expressed by the total number of web pages each user browsed throughout the tracking period.

\textit{RQ2: temporal patterns of web search.} To examine temporal patterns of web search and their differences by country and demographics, we have calculated the frequency of web search use 
by day of the week and time of day (morning = 6am to 12pm; afternoon = 12pm to 6pm; evening = 6pm to midnight; night = midnight to 6am). We then compared the results in terms of the patterns that emerged.

\textit{RQ3: search engine preferences.} To assess the differences in the choices of specific search engines by demographics, we calculated descriptive statistics across engines. First, we calculated the share of each search engine in overall web search traffic in our sample. Then, for the search engines that accounted for at least 1\% of search traffic in either of the two (German and Swiss) samples, we calculated the share of users in each demographic (gender; age) group that used the engine at least once during the observation period. Then, among those users who used an engine at least once, we calculated the average share of search traffic each user (disaggregated by gender and age groups) devoted to a specific engine to assess the strength of users' preferences towards specific engines. As Google, unsurprisingly, emerged as a
dominant engine in all of the cases, we decided to examine whether demographic and socio-economic characteristics of users are associated with their likelihood to "ditch" Google in favor of one of other
engines. For that, we relied on regression analysis using zero-inflated Poisson models due to the nature of the data (i.e., most users recorded 0 visits to engines other than Google). As a dependent variable, we used participants' share of visits to engines other than Google in their overall search traffic (varying from 0 to 100), and as predictors we used socio-economic and demographic characteristics, controlling for the total number of web pages visited.

\textit{RQ4: result ranking and clicking behavior.} To establish the association between result ranking and clicking behavior, we have extracted links for
organic text results from the HTML files corresponding to 
pages browsed by the users for Google and Bing. We focused on these two engines due to their relative prevalence in users' search (see Results, subsection 1). We extracted the links in the same order as they appeared in users' browsers. Then, we matched this data to the URLs accessed by users after each visit to a search engine to infer which of the URLs displayed in search results a user clicked on. Based on this inference, we calculated summary statistics about the share of clicks different search pages and differently ranked results received. Then, to infer whether certain demographic or socio-economic characteristics contribute to users' likelihood to click on higher- or lower-ranked results, we performed regression analysis using an average ranking of a Google search result a user clicked on as a dependent variable, their demographic and socio-economic characteristics as predictors, and total number of web pages visited by them as control. We focused here exclusively on Google due to its dominance in users' web search - thus, running the analysis on Google, unlike on other engines, allowed us to preserve for the analysis the number of participants that is high enough to be 
representative of the general sample.

\section{Results}
\subsection{Frequency of search engine use by demographics}

\begin{table*}[t]
\begin{tabular}{|l|l|l|l|l|l|l|l|}
\hline
\textbf{Country sample}              & \textbf{Variable}                                             & \textbf{Overall} & \textbf{Women}  & \textbf{Men}    & \textbf{Age 18-34} & \textbf{Age 35-54} & \textbf{Age 55+} \\ \hline
\multirow{2}{*}{Germany}     & Share of search in browsing & 10.4\%  & 10.6\% & 10.2\% & 12\%      & 10.2\%    & 10\%                    \\ \cline{2-8} 
                             & Average number of searches daily           & 8.84    & 7.71   & 9.75   & 9.52      & 9.72      & 7.96                    \\ \hline
\multirow{2}{*}{Switzerland} & Share of search in browsing & 15.5\%  & 17.1\% & 14.2\% & 18.6\%    & 14.6\%    & 12.8\%                  \\ \cline{2-8} 
                             & Average number of searches daily           & 8.76    & 8.51   & 8.95   & 11.8      & 7.65      & 6.33                    \\ \hline
\end{tabular}
\caption{Average ratio of web search visits to the overall number of visits tracked and average number of searches executed daily per user, differentiated by country subsamples and demographic groups.}
\label{summarystats}
\end{table*}

On average, participants used text search 8.8 times per day (see Table \ref{summarystats}). There were no major differences in the number of average daily searches between the subsamples from
the two countries: in Switzerland, users engaged in web search on average 8.76 times per day, while in Germany 8.84 times per day. However, there were differences in the
ratio of web search visits to the overall number of visits tracked. In the overall sample, on average each user had 13\% of their total tracked browsing devoted to web search. In the German sample this number was 10.4\%, while in the Swiss one - 15.5\%. Thus though the users from both countries executed similar numbers of searches on a daily basis, in the Swiss sample web search accounted for a higher share of total internet browsing. This is in line with the discrepancy in the average number of pages browsed in total in each sample: 2322.1 pages in Switzerland versus 3833.7 in Germany.

In Table \ref{summarystats} we present summary statistics for the average share of search in web browsing per user and average number of searches executed daily per user disaggregated by demographic groups with respect to gender and age. Though there are apparent gender-based discrepancies with regard to the average number of searches executed daily per user in Germany and with regard to the share of search in overall browsing in Switzerland, in regression analysis gender is not a significant predictor in either sample. The only significant predictor of either of the aforementioned variables is age. In Switzerland, it is negatively associated with both the average daily number of searches and the share of search in total browsing. In Germany, the observed relationship is significant only for the former variable but not for the latter. Detailed regression outputs are presented in Tables \ref{daily_regression} and \ref{shareregression}.

\begin{table}
\begin{center}
\resizebox{.99\columnwidth}{!}{
\begin{tabular}{l D{)}{)}{9)3} D{)}{)}{9)3}}
\toprule
 & \multicolumn{1}{c}{Germany} & \multicolumn{1}{c}{Switzerland} \\
\midrule
Intercept                     & 9.44 \; (2.15)^{***} & 8.01 \; (1.63)^{***}  \\
Gender (female)               & -0.73 \; (0.75)      & 0.76 \; (0.68)        \\
Age                           & -0.07 \; (0.03)^{**} & -0.11 \; (0.02)^{***} \\
Education level               & -0.06 \; (0.56)      & 0.55 \; (0.62)        \\
Income                        & 0.01 \; (0.31)       & 0.13 \; (0.23)        \\
Total number of pages browsed & 0.00 \; (0.00)^{***} & 0.00 \; (0.00)^{***}  \\
\midrule
AIC                           & 3829.88              & 3543.97               \\
Log Likelihood                & -1907.94             & -1764.99              \\
Deviance                      & 36641.08             & 26428.36              \\
Num. obs.                     & 541                  & 522                   \\
\bottomrule
\multicolumn{3}{l}{\scriptsize{$^{***}p<0.001$; $^{**}p<0.01$; $^{*}p<0.05$}}
\end{tabular}
}
\caption{Regression model output, predicting average daily number of searches executed by each user}
\label{daily_regression}
\end{center}
\end{table}

\begin{table}
\begin{center}
\resizebox{.99\columnwidth}{!}{
\begin{tabular}{l D{)}{)}{9)3} D{)}{)}{9)3}}
\toprule
 & \multicolumn{1}{c}{Germany} & \multicolumn{1}{c}{Switzerland} \\
\midrule
Intercept                     & 0.12 \; (0.02)^{***}  & 0.24 \; (0.02)^{***}  \\
Gender (female)               & -0.00 \; (0.01)       & 0.02 \; (0.01)        \\
Age                           & -0.00 \; (0.00)       & -0.00 \; (0.00)^{***} \\
Education level               & 0.00 \; (0.01)        & 0.00 \; (0.01)        \\
Income                        & 0.00 \; (0.00)        & -0.00 \; (0.00)       \\
Total number of pages browsed & -0.00 \; (0.00)^{***} & -0.00 \; (0.00)^{***} \\
\midrule
AIC                           & -1089.05              & -823.34               \\
Log Likelihood                & 551.52                & 418.67                \\
Deviance                      & 4.12                  & 6.15                  \\
Num. obs.                     & 541                   & 522                   \\
\bottomrule
\multicolumn{3}{l}{\scriptsize{$^{***}p<0.001$; $^{**}p<0.01$; $^{*}p<0.05$}}
\end{tabular}
}
\caption{Regression model output, predicting average share of text search in a user's browsing}
\label{shareregression}
\end{center}
\end{table}

\subsection{Temporal patterns of web search}
\begin{figure*}[t]
\centering
\includegraphics[width=0.8\textwidth]{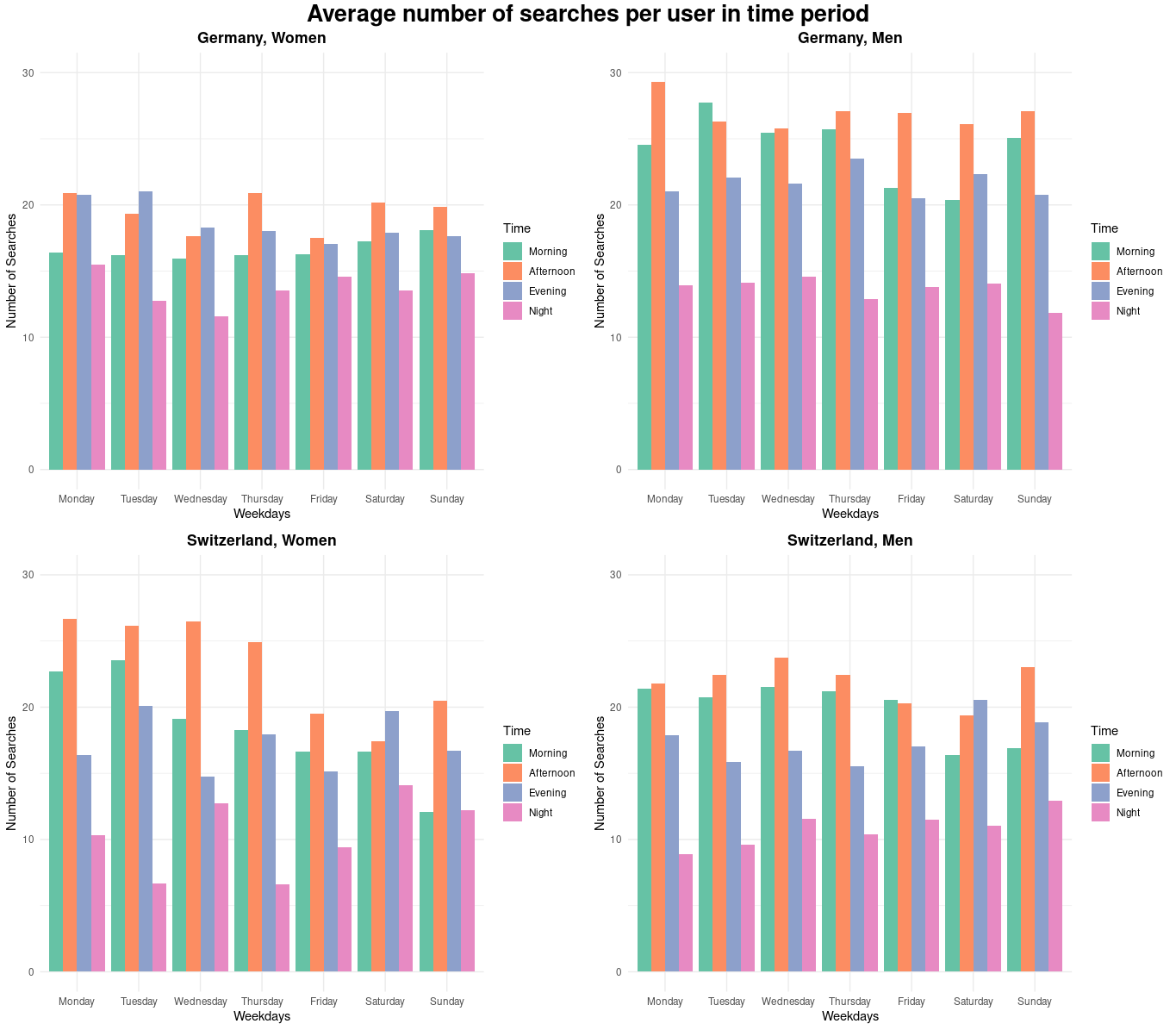} 
\caption{Average number of searches executed per user in weekday and time of day periods, disaggregated by country and gender.}
\label{temporal}
\end{figure*}
The analysis of temporal patterns of web search reveals major differences with regard to when users of different gender tend to use search engines. The average numbers of searches executed per user on each weekday and time of the day periods are presented in Fig.~\ref{temporal}.

In both, Switzerland and Germany, men's search patterns are stable throughout the week, with the most active search times being morning and afternoon. In the evenings, searching is less prevalent than in the mornings and afternoons, followed by a major drop in searching at night. Women's search patterns, however, are different. In Germany, they are similarly to men's consistent throughout the week but women tend to search way less than men in the mornings and afternoons, slightly less in the evenings, and with the same intensity
in the night. The differences in German women's search activity by the time of the day are way less drastic than in men's search activity. In Switzerland, on the other hand, time-of-day-based differences in women's search activity are similar to those observed for men. However, Swiss women's searches in the mornings and afternoons are unevenly distributed throughout the week: from Monday to Thursday women actively use web search in the afternoons, but from Friday to Sunday their search activity in the afternoon is reduced compared to the first part of the week. Morning search activity among Swiss women decreases progressively from the start to the end of the week. 

The observed differences among women and men in both countries are the most pronounced among younger (18-45 years old) part of the populations, suggesting that they might have to do with the life circumstances of younger women such as, potentially, child-bearing and child-caring duties. However, based on the available data, it is hardly possible to verify this interpretation. Regardless of the reasons behind the observed discrepancies, our findings once again highlight the necessity of disaggregating data about information behaviour by gender \cite{perez_invisible_2019}
to grasp the behavioral patterns of all parts of the population properly.

\subsection{Usage of specific search engines}

\begin{table}[t]
\resizebox{.99\columnwidth}{!}{
\begin{tabular}{|l|l|l|l|l|l|}
\hline
\textbf{Country}     & \textbf{Google}  & \textbf{Bing} & \textbf{Ecosia}   & \textbf{DDG} & \textbf{Other (total)} \\ \hline
Germany     & 88\%   & 7.4\%  & 2.5\% & 0.5\%      & 1.6\%         \\ \hline
Switzerland & 94.4\% & 1.5\%    & 2\% & 1.3\%      & 0.8\%         \\ \hline
\end{tabular}
}
\caption{Share of searches within a specific search engine among all searches from all users by the country subsample. The numbers are reported only for specific search engines for which the share is above 1\% in one of the subsamples. The share of all other search engines is aggregated as "Other".}
\label{searchgeneraltraffic}
\end{table}

\begin{table*}[t]
\centering
\begin{tabular}{|l|l|l|l|l|l|l|l|}
\hline
\textbf{Country sample}               & \textbf{Engine} & \textbf{Overall} & \textbf{Women}  & \textbf{Men}    & \textbf{Age 18-34} & \textbf{Age 35-54} & \textbf{Age 55+} \\ \hline
\multirow{4}{*}{Germany}     & Google & 96.6\%  & 96.4\% & 96.8\% & 96.8\%    & 98.3\%    & 94.8\%                  \\ \cline{2-8} 
                             & Bing   & 17.7\%  & 15.6\% & 19.5\% & 11.8\%    & 18.8\%    & 19.1\%                  \\ \cline{2-8} 
                             & Ecosia & 4.5\%   & 4\%    & 4.9\%  & 4.3\%     & 4.3\%     & 4.9\%                   \\ \cline{2-8} 
                             & DDG    & 1.8\%   & 2\%    & 1.6\%  & 0\%       & 2.1\%     & 2.8\%                   \\ \hline
\multirow{4}{*}{Switzerland} & Google & 97.9\%  & 98\%   & 97.8\% & 97.6\%    & 98.9\%    & 97.1\%                  \\ \cline{2-8} 
                             & Bing   & 10.7\%  & 8.9\%  & 12\%   & 11.2\%    & 10.4\%    & 10.3\%                  \\ \cline{2-8} 
                             & Ecosia & 3.2\%   & 4.5\%  & 2.2\%  & 4.9\%     & 1.1\%     & 3.4\%                   \\ \cline{2-8} 
                             & DDG    & 3.4\%   & 2.4\%  & 4.1\%  & 3.4\%     & 2.8\%     & 4\%                     \\ \hline
\end{tabular}
\caption{Share of participants who used a specific search engine at least once by country, gender and age group.}
\label{usedonce}
\end{table*}

\begin{table}[]
\resizebox{.99\columnwidth}{!}{
\begin{tabular}{|l|l|l|l|l|l|l|}
\hline
\textbf{Country} & \textbf{Overall} & \textbf{Women}  & \textbf{Men}    & \textbf{Age 18-34} & \textbf{Age 35-54} & \textbf{Age 55+} \\ \hline
Germany        & 26.9\%  & 21.6\% & 31.2\% & 16\%      & 29.9\%    & 28.3\%  \\ \hline
Switzerland    & 23.6\%  & 21.5\% & 25.2\% & 20.4\%    & 22.5\%    & 23.6\%  \\ \hline
\end{tabular}
}
\caption{Share of users in each demographic group who used more than one search engine.}
\label{morethanonce}
\end{table}

\begin{table*}[t]
\centering
\begin{tabular}{|l|l|l|l|l|l|l|l|}
\hline
\textbf{Country sample}               & \textbf{Engine} & \textbf{Overall} & \textbf{Women}  & \textbf{Men}    & \textbf{Age 18-34} & \textbf{Age 35-54} & \textbf{Age 55+} \\ \hline
\multirow{4}{*}{Germany}     & Google & 90.2\%  & 91.4\% & 89.2\% & 93.8\%    & 90.5\%    & 88.3\%  \\ \cline{2-8} 
                             & Bing   & 37.1\%  & 42.5\% & 33.6\% & 48.1\%    & 26.9\%    & 44.5\%  \\ \cline{2-8} 
                             & Ecosia & 61.1\%  & 66.9\% & 57.3\% & 75.9\%    & 54.6\%    & 61.6\%  \\ \cline{2-8} 
                             & DDG    & 59.8\%  & 61.9\% & 56.7\% & ---       & 63\%      & 56.1\%  \\ \hline
\multirow{4}{*}{Switzerland} & Google & 94.1\%  & 94.9\% & 93.5\% & 94.4\%    & 95.9\%    & 91.8\%  \\ \cline{2-8} 
                             & Bing   & 28\%    & 30.3\% & 27\%   & 21.8\%    & 17\%      & 48.3\%  \\ \cline{2-8} 
                             & Ecosia & 55.7\%  & 52.6\% & 60.6\% & 58.9\%    & 58\%      & 49.6\%  \\ \cline{2-8} 
                             & DDG    & 52\%    & 60.1\% & 48.3\% & 62.1\%    & 49.4\%    & 43.8\%  \\ \hline
\end{tabular}
\caption{Mean share of usage of specific search engines in all search visits per user, among participants who used a specific engine at least once.}
\label{searchshares_once}
\end{table*}

\begin{table*}[t]
\begin{center}
\begin{tabular}{l D{)}{)}{9)3} D{)}{)}{9)3} D{)}{)}{9)3}}
\toprule
 & \multicolumn{1}{c}{Bing} & \multicolumn{1}{c}{DuckDuckGo} & \multicolumn{1}{c}{Ecosia} \\
\midrule
Count model: Intercept                     & 3.73 \; (0.17)^{***}  & 4.81 \; (0.23)^{***}  & 4.19 \; (0.21)^{***}  \\
Count model: Gender (female)               & 0.07 \; (0.06)        & 0.15 \; (0.09)        & 0.69 \; (0.11)^{***}  \\
Count model: Age                           & 0.01 \; (0.00)^{***}  & 0.00 \; (0.00)        & -0.01 \; (0.00)^{***} \\
Count model: Education level               & -0.14 \; (0.05)^{**}  & -0.02 \; (0.07)       & -0.33 \; (0.08)^{***} \\
Count model: Income                        & -0.05 \; (0.02)^{*}   & -0.17 \; (0.03)^{***} & 0.28 \; (0.05)^{***}  \\
Count model: Total number of searches & -0.00 \; (0.00)^{***} & -0.00 \; (0.00)^{***} & -0.00 \; (0.00)^{***} \\
Zero model: Intercept                      & 3.22 \; (0.90)^{***}  & 2.70 \; (1.31)^{*}    & 6.14 \; (1.62)^{***}  \\
Zero model: Gender (female)                & -0.15 \; (0.36)       & 0.83 \; (0.58)        & -0.25 \; (0.54)       \\
Zero model: Age                            & -0.01 \; (0.01)       & 0.00 \; (0.02)        & 0.00 \; (0.02)        \\
Zero model: Education level                & 0.33 \; (0.32)        & -0.52 \; (0.52)       & -1.65 \; (0.58)^{**}  \\
Zero model: Income                         & -0.29 \; (0.12)^{*}   & 0.46 \; (0.20)^{*}    & 0.43 \; (0.19)^{*}    \\
Zero model: Total number of searches  & -0.00 \; (0.00)       & 0.00 \; (0.00)        & 0.00 \; (0.00)        \\
\midrule
AIC                                        & 1829.83               & 615.75                & 657.19                \\
Log Likelihood                             & -902.92               & -295.88               & -316.59               \\
Num. obs.                                  & 522                   & 522                   & 522                   \\
\bottomrule
\multicolumn{4}{l}{\scriptsize{$^{***}p<0.001$; $^{**}p<0.01$; $^{*}p<0.05$}}
\end{tabular}
\caption{Zero-inflated model outputs, predicting the usage of specific most popular search engines except Google, Switzerland.}
\label{zeroinf_ch_engines}
\end{center}
\end{table*}

\begin{table*}[t]
\begin{center}
\begin{tabular}{l D{)}{)}{9)3} D{)}{)}{9)3} D{)}{)}{9)3}}
\toprule
 & \multicolumn{1}{c}{Bing} & \multicolumn{1}{c}{DuckDuckGo} & \multicolumn{1}{c}{Ecosia} \\
\midrule
Count model: Intercept                     & 3.30 \; (0.11)^{***}  & 5.54 \; (0.51)^{***}  & 3.62 \; (0.17)^{***}  \\
Count model: Gender (female)               & 0.24 \; (0.04)^{***}  & -0.63 \; (0.12)^{***} & 0.07 \; (0.06)        \\
Count model: Age                           & 0.01 \; (0.00)^{***}  & -0.01 \; (0.01)       & -0.01 \; (0.00)^{***} \\
Count model: Education level               & -0.24 \; (0.03)^{***} & -0.06 \; (0.12)       & 0.47 \; (0.06)^{***}  \\
Count model: Income                        & 0.12 \; (0.02)^{***}  & -0.25 \; (0.04)^{***} & -0.05 \; (0.02)^{*}   \\
Count model: Total number of searches & 0.00 \; (0.00)^{***}  & 0.00 \; (0.00)^{***}  & 0.00 \; (0.00)        \\
Zero model: Intercept                      & 1.01 \; (0.74)        & 4.30 \; (2.08)^{*}    & 2.44 \; (1.32)        \\
Zero model: Gender (female)                & 0.14 \; (0.26)        & -0.26 \; (0.71)       & 0.43 \; (0.46)        \\
Zero model: Age                            & -0.01 \; (0.01)       & -0.00 \; (0.03)       & -0.02 \; (0.02)       \\
Zero model: Education level                & 0.27 \; (0.19)        & -0.37 \; (0.56)       & -0.10 \; (0.35)       \\
Zero model: Income                         & 0.13 \; (0.11)        & 0.15 \; (0.29)        & 0.50 \; (0.19)^{**}   \\
Zero model: Total number of searches  & -0.00 \; (0.00)       & 0.00 \; (0.00)        & 0.00 \; (0.00)        \\
\midrule
AIC                                        & 3485.40               & 331.98                & 875.53                \\
Log Likelihood                             & -1730.70              & -153.99               & -425.76               \\
Num. obs.                                  & 541                   & 541                   & 541                   \\
\bottomrule
\multicolumn{4}{l}{\scriptsize{$^{***}p<0.001$; $^{**}p<0.01$; $^{*}p<0.05$}}
\end{tabular}
\caption{Zero-inflated model outputs, predicting the usage of specific most popular search engines except Google, Germany.}
\label{zeroinf_de_engines}
\end{center}
\end{table*}

Google largely dominates search traffic in both country subsamples, being slightly more prevalent
in Switzerland than in Germany, as demonstrated in Table \ref{searchgeneraltraffic}. In Germany, it is followed by Bing with the latter accounting for 7.4\% of all traffic - substantially lower than Google but more than 3 times higher than all other engines in either of the country subsamples. In Switzerland, Bing is less prominent, with Ecosia being the second most popular engine that accounts for 
\% of search traffic. The only other engine that received at least 1\% of the traffic in either of the samples is DuckDuckGo (DDG) with 1.3\% of traffic in Switzerland. These findings are in line with the reports made by companies monitoring global search traffic on country-level such as Statcounter \cite{noauthor_desktop_nodate}.

In Table \ref{usedonce} we report the observations on the share of participants who used each search engine at least once disaggregated by demographics. In light of the already stated observations about the share of search traffic, it comes as no surprise that Google was used at least once by 
almost all participants 
The second most popular engine by the share of users who turned to it at least once is Bing. It was used at least once by 17.7\% of German participants and 10.7\% of Swiss participants, with the share of men turning to Bing being higher than the share of women in both cases. The patterns of Bing usage by age groups, however, vary between the two countries with its prevalence being higher among elder users in Germany and among younger users in Switzerland. The other engines were used at least once by a marginal share of users 
- less than 5\% across both subsamples and all demographic groups. We observe no consistent gender/age patterns with regard to Ecosia and DuckDuckGo usage across the two countries.

In Table \ref{searchshares_once} we report average shares of search traffic via a given search engine per user among participants who used the engine at least once, and in Table \ref{morethanonce} the share of users who used more than one search engine. This way we can assess how "partisan" participants are in terms of their search engine preferences - i.e., whether participants from different demographic groups tend to search within one search engine almost exclusively or engage in search across different engines, and whether this partisanship varies from one engine to another. 

In both samples, around a quarter of all participants used more than one search engine; the share of such participants tends to be higher among men and older users in both cases (Table \ref{morethanonce}). We observe that Google tends to be the default search engine for all demographic groups in both countries, with participants who used it at least once directing around 90\% of their search traffic there. The participants were less "partisan" in their usage of other search engines, with differences in the levels of "partisanship" between Bing vs Ecosia and DuckDuckGo. Though Bing has been used at least once by a larger proportion of participants than the other engines (Table \ref{usedonce}), most of the demographic groups tend to direct only around 30-40\% of their search traffic through this engine, thus indicating the absence of a strong preference for it. Those who used Ecosia or DuckDuckGo tend to be more "partisan" in their preference for these engines, directing a 50-60\% of search through them, though the strength of the preference is clearly way lower than that of Google users.

In Tables \ref{zeroinf_ch_engines} and \ref{zeroinf_de_engines} we report the results of regression analysis conducted to establish which - if any - demographic and socio-economic characteristics are associated with a higher likelihood of using non-dominant  - that is, not Google - search engines.

In both countries, we observe no relationship between gender and age and the one's likelihood to use a search engine other than Google (zero model outputs). However, in Switzerland income is negatively associated with the usage of Bing, while its the association with the usage of Ecosia and DuckDuckGo is positive. In Germany, income is positively associated only with the usage of Ecosia, and no other demographic or socio-economic factors are significantly associated with the usage of specific
search engines. In Switzerland, the usage of Ecosia is also negatively associated with the participants' level of education.

When it comes to the prevalence of use of each of the search engines among the participants that used each engine at least once (count model outputs), a lot of significant relationships emerge. Gender (being female) is positively associated with the usage of Ecosia in Switzerland and the usage of Bing in Germany, and negatively associated with the usage of DuckDuckGo in Germany. In both countries, there is a positive association between age and the usage of Bing and a negative association between age and the usage of Ecosia. The relationship between the usage of DuckDuckGo and age is significant (and negative) only in Switzerland. Education level has significant relationships only with the usage of Ecosia and Bing both countries. For Bing, the relationship is negative in both cases, and for Ecosia positive in Germany and negative in Switzerland. Finally, income in Switzerland is negatively related with the usage of Bing and DuckDuckGo and positively with the usage of Ecosia. In Germany, the directionality of the relationship is the same for DuckDuckGo but reversed for Ecosia and Bing.

Overall, these observations suggest that there are country-level differences in the users' preferences towards different search engines, including the relation between such preferences and users' demographics and socio-economic status. Importantly, income was significant more often than other variables, thus suggesting that participants with different economic status tend to have a preference for different engines.

\subsection{Ranking of search results and user clicking behavior}
We have examined users' clicking behavior on Google and Bing - the two engines which were visited most frequently by the users in our sample. We found that on both engines participants clicked disproportionately more on top results. On Google, 97.11\% percent of all clicks were associated with the results displayed on the first page; on Bing this number was even higher - 99.49\%. 

Even within the first page, users' clicks are distributed unequally. On Google, 51.3\% of all clicks were associated with the very first result, followed by 15.68\% of clicks on the second result, 9.23\% on the third, 5.93\% on the fourth, 4.21\% on the fifth. As such, top-5 
search results accounted for over 86\% of all clicks 
on Google. On Bing, the corresponding distribution of clicks for the top-5 results is as follows: 52.18\%; 19.72\%; 9.24\%; 7.35\%; 3.57\%. Thus, on Bing top-5 results received 
around 92\% of all clicks. We did not observe major differences in the users' clicking behavior between the two engines, suggesting that it does not depend on the engine, at least when both engines have a similar (i.e., results presented in a ranked list) interface.

We found no significant relationship between (Google) users' likelihood of clicking on higher- or lower-ranked results and their demographic or socio-economic characteristics in Germany (see Table \ref{ranking_google}). In Switzerland, only age emerged as a significant variable, though its effect size is small. This suggests that the relationship between search results ranking and users' clicking behavior is largely independent from users' demographic and socio-economic characteristics. However, there seems to be a difference in the behavior of the users from two countries as in the Swiss subsample tendency to click on higher-ranked results is even more pronounced than in the German one. The mean of the average ranking of the results clicked per user for Switzerland is 2.49 (median=2.3), and in Germany the corresponding values are 3.13 and 2.55.
\begin{table}
\begin{center}
\resizebox{.99\columnwidth}{!}{
\begin{tabular}{l D{)}{)}{8)1} D{)}{)}{9)3}}
\toprule
 & \multicolumn{1}{c}{Germany} & \multicolumn{1}{c}{Switzerland} \\
\midrule
Intercept                     & 1.29 \; (0.97)     & 0.90 \; (0.46)       \\
Gender (female)               & 0.03 \; (0.34)     & -0.04 \; (0.18)      \\
Age                           & 0.02 \; (0.01)     & 0.02 \; (0.01)^{***} \\
Education level               & 0.13 \; (0.25)     & 0.08 \; (0.17)       \\
Income                        & 0.04 \; (0.14)     & 0.07 \; (0.06)       \\
Share of search in browsing   & 5.69 \; (6.72)     & 5.93 \; (3.11)       \\
Total number of pages browsed & 0.00 \; (0.00)^{*} & 0.00 \; (0.00)^{**}  \\
\midrule
AIC                           & 2410.91            & 1880.25              \\
Log Likelihood                & -1197.45           & -932.13              \\
Deviance                      & 5243.70            & 1519.78              \\
Num. obs.                     & 453                & 463                  \\
\bottomrule
\multicolumn{3}{l}{\scriptsize{$^{***}p<0.001$; $^{**}p<0.01$; $^{*}p<0.05$}}
\end{tabular}
}
\caption{Regression model output, dependent variable is the average ranking of Google search results clicked on by the user}
\label{ranking_google}
\end{center}
\end{table}

\section{Limitations}
Our study has two major limitations. First, we examine only desktop-based browsing behaviour because mobile-based tracking is notoriously complex to implement, especially in a way
that would make mobile data comparable with the 
the desktop one 
\cite{christner_automated_2021}. This is an important limitation given that mobile devices account for roughly a half of the global internet traffic. Also, given the possibility that users' desktop- and mobile-based behaviours might differ, our findings have to be interpreted only in the context of desktop browsing. We also suggest that future work focusing on mobile browsing specifically is necessary to gain important insight about the differences in desktop- and mobile-based searching behaviours. Another limitation is that our data collection happened to take place in the spring of 2020 - the beginning of the COVID-19 pandemic - and thus largely coincided with the lockdowns in both, Germany and Switzerland. In light of this, our findings, especially those related to the temporal patterns of web search, need to be interpreted with caution, because it is difficult to evaluate whether and how much users' search behaviours during the lockdowns might have been different from those in routine times.

\section{Discussion}
Our analysis shows substantial differences across countries in all four aspects of web search behaviour we examined. This highlights the need for more comparative research on web search behaviour. Since 
behaviours of users from Germany and Switzerland, two geographically and, in part, culturally proximate countries, are vastly different, it is reasonable to assume that the behaviours of those in other countries are different as well. Thus, like with other online phenomena \cite{krishnan_determinants_2017,urman_context_2019,mahmood_-line_2004}, the context of the country in which web-based behaviour is studied needs to be accounted for, and generalizations from single-country samples to the global populations should be avoided.

The fact that web search accounts for a rather high share of desktop browsing (~13\%), highlights the importance of search engines for the public. Similarly, 
the fact that top-5 results attract ~90\% of all clicks and more than 
97\% of search visits
do not go beyond page 1 of search outputs, underscores the influence search rankings have on user information consumption. Given that search engines' 
retrieval and ranking algorithms are usually obscure
and the outputs
are heavily affected by personalization \cite{hannak_measuring_2013,kliman-silver_location_2015,robertson_auditing_2018} and randomization \cite{makhortykh_how_2020,urman_matter_2021}, algorithmic auditing studies, with a particular focus on the ranking of top search results, are necessary to understand how specific factors affect the search outputs and the quality of information consumed by the users.

As web search algorithms tend to optimize the results based on user behaviour \cite{agichtein_improving_2006}, demographic and socio-economic status-based differences in the ways users search the web can have important implications for the selection of information that users access. For instance, if a search engine, especially that without explicit personalization such as DuckDuckGo, is used disproportionately more by users with certain characteristics (i.e., male users and/or those with lower income, see Tables \ref{zeroinf_ch_engines}, \ref{zeroinf_de_engines}), the search results will be tailored to the preferences of users with such characteristics. In extreme cases, this can lead to systematic biases in search results - e.g., the perpetuation of "male gaze" in results concerning the representation of women \cite{noble_algorithms_2018}. 

\bibliography{Paper}

\end{document}